\documentclass[preprint,12pt, a4paper]{elsarticle}

\usepackage{amssymb}
\usepackage{hyperref}
\usepackage{xcolor}
\journal{SoftwareX}

\begin{document}
\renewcommand{\labelenumii}{\arabic{enumi}.\arabic{enumii}}

\begin{frontmatter}



\title{VR-IPS: Virtual Reality Tool for Photic Stimulation in Photosensitive Diagnosis}


\author{Fernando Moncada Martins$^{1,3}$, Daniel P{\'e}rez Pr{\'a}danos$^{2}$, Angel Rio-Alvarez$^{1,3}$, MG Cano-Celestino$^{6}$, Mar{\'i}a Antonia Guti{\'e}rrez$^{4}$, Pablo Calvo Calleja$^{4}$, D. Kasteleijn-Nolst Trenite$^{5}$, and V{\'i}ctor M. Gonz{\'a}lez$^{2,3}$}

\address{$^{1}$ Computer Science Department, University of Oviedo, Spain \\
$^{2}$ Electrical Engineering Department, University of Oviedo, Spain \\
$^{3}$ Biomedical Engineering Center (BME), University of Oviedo, Spain \\
$^{4}$ Neurophysiology Service, Cabue{\~n}es University Hospital, Spain \\
$^{5}$ Nesmos Department, Faculty of Medicine and Psychology, Sapienza University, Rome, Italy\\
$^{6}$ Centro de Ciencias Cognitivas, Univresidad Autónoma San Luis Potosí, México}

\begin{abstract}
Photosensitivity is a neurological condition in which the brain generates epileptiform activity in response to visual stimuli. The standardized clinical diagnosis procedure, named Intermitent Photic Stimulation (IPS), involves exposing patients to a white flashing light at different frequencies to provoke this reaction. However, clinical neurophysiologists report that this protocol is insufficient, leading to underdiagnosis. VR-IPS is a flexible visual stimulation tool that extends conventional flashing-light stimulation by incorporating configurable color lights, static and oscillatory motion patterns, and photoprovocative videos. The system comprises a stimulation module that can run on Virtual Reality headsets or on standard monitors, together with a control module that allows clinicians to create, select, and configure visual stimulus, while controlling the whole stimulation procedure.
\end{abstract}


\end{frontmatter}


\section*{Required Metadata}

\section*{Current code version}

\begin{table}[!h]
\begin{tabular}{|l|p{6.5cm}|p{6.5cm}|}
\hline
\textbf{Nr.} & \textbf{Code metadata description} & \textbf{Please fill in this column} \\
\hline
C1 & Current code version & v1.0 \\
\hline
C2 & Permanent GitHub link to code/repository used for this code version & \colorbox{lime}{PROYECTO PUBLICO} \url{https://github.com/AI-Biomedical-Engineering/vrphotosense2} \\
\hline
C3 & Legal Code License & GNU General Public License v3.0 (GPL-3.0) \\
\hline
C4 & Code versioning system used & Git \\
\hline
C5 & Software code languages, tools, and services used & JavaScript, C\#, CSS, HTML \\
\hline
C6 & Compilation requirements, operating environments \& dependencies & Microsoft Windows or Linux OS (for Desktop version), Oculus Quest 2 headset (for VR version), Android smartphone (for Smartphone version) \\
\hline
C7 & If available Link to developer documentation/manual & Included in Repository \\
\hline
C8 & Support email for questions & moncadafernando@uniovi.es \\
\hline
C9 & Permanent link to Reproducible Capsule & \url{https://visir.edv.uniovi.es} \\
\hline
\end{tabular}
\caption{Code metadata (mandatory)}
\label{} 
\end{table}


\textbf{Main text}

\begin{enumerate}

\item Motivation and significance 

Photosensitivity is a neurological condition in which the brain generates epileptiform electrical activity as a reaction to certain visual stimuli, such as flashing lights or photoprovocative visual patterns. The standardized clinical diagnostic procedure, known as \textit{Intermittent Photic Stimulation} (IPS), comprises a white LED light positioned approximately 30cm from the patient's eyes. The light flashes at predefined frequencies to elicit epileptiform discharges referred to as Photoparoxysmal Responses (PPR), which constitute the primary electroencephalographic (EEG) biomarker used to diagnose photosensitivity. 

Epidemiological studies indicate that the prevalence of photosensitivity is extremely low \cite{Dorothee1989,Fisher2022}. 
Despite its established clinical relevance, neurophysiologists recognize significant limitations in conventional IPS. Photosensitive epilepsy patients can complain about their sensitivity to a variety of visual stimuli. Just conventional flashing white light does not cover this variety; the restricted variety of available visual stimuli does not adequately represent the diversity of photoprovocative conditions encountered in everyday life. Consequently, some patients who show clear symptoms do not exhibit any PPR during the IPS procedure, leading to underdiagnosis or missed diagnoses.

To address these limitations, we designed a clinical visual stimulation tool designed to provide richer and more flexible photic stimulation scenarios during photosensitivity assessments. Developed in close collaboration with expert clinical neurophysiologists, it extends conventional IPS by supporting configurable color stimuli, static and dynamic patterns, and photoprovocative videos on both Virtual Reality (VR) and conventional monitor-based devices. It provides clinicians with a dedicated control module to create and configure as many stimuli and stimulation sequences as desired, and to control the stimulation procedure. \\


\item Software description 

VR-IPS enables clinical neurophysiologists to create, configure, and deliver a wide variety of photoprovocative visual stimuli, enabling a more comprehensive characterization of each patient's photosensitive profile. 

The software consists of two modules: i) the \textbf{Control Module} (\textbf{CM}), used to create and to configure the stimuli and to control the stimulation process; and ii) the \textbf{Stimulation Module} (\textbf{SM}), which screens the selected visual stimuli to the patient. 

Both modules are accessible through the \textit{Permanent link to Reproducible Capsule}. It first launches the \textbf{CM}, from which the appropriate version of the \textbf{SM} can be downloaded for the target stimulation hardware. Fig.\ref{fig:system} illustrates an overview of the complete system architecture. \\

\begin{figure}
    \centering
    \includegraphics[width=0.85\linewidth]{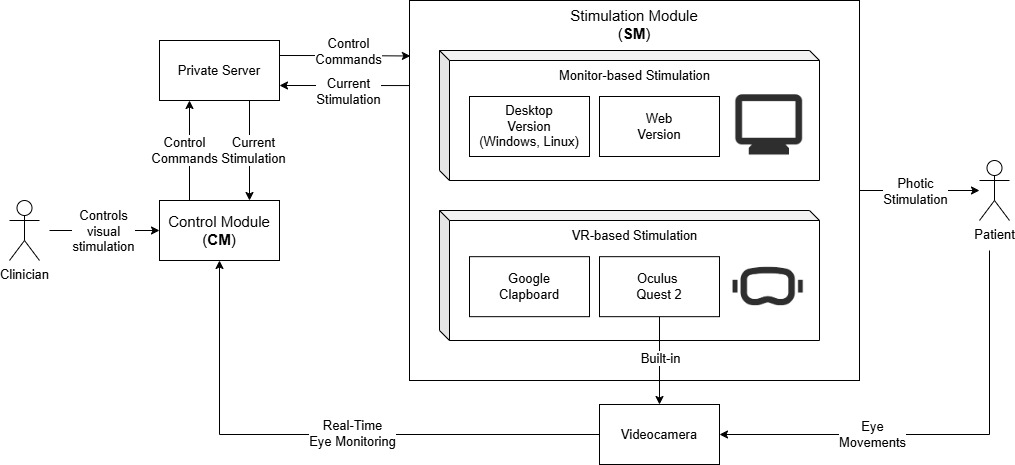}
    \caption{VRPhotosense2 architecture.}
    \label{fig:system}
\end{figure}


\begin{enumerate}

\item Stimulation Module 

Three versions of the \textbf{SM} are available: for VR headsets, for smartphones, and for conventional monitors. Monitor-based stimulation provides the simplest and most widely accessible deployment option, which is already used in other clinical assessments. However, conventional displays inherently limit the effectiveness of certain stimuli. More immersive and realistic scenarios, which may be relevant for eliciting PPR discharges in some patients, can only be fully achieved using VR-based devices. As an intermediate solution, cardboard-style headsets offer a VR-like experience by placing a smartphone at a viewing distance comparable to that of dedicated VR headsets.

The monitor-based version is available both as a web application, ---requiring no download nor installation---, and as a standalone desktop application for Windows and Linux that can be run without installation. The VR and smartphone versions are distributed as \textit{APK} packages that must be downloaded and installed on the target device before use.

Once running, the \textbf{SM} can establish a connection with the \textbf{CM} through the Private Server where it is deployed. During the stimulation session, the \textbf{CM} transmits the selected stimulus, which the \textbf{SM} plays on the visualization device. \\

\item Eye Monitoring 

During the design process, clinical neurophysiologists pointed out the necessity for continuous monitoring of the patient's eyes as a key factor when using VR-based stimulation. The selected VR device for this development was the Oculus Quest 2, which does not include built-in eye cameras. To overcome this limitation, two miniature cameras were integrated into the device and pointed toward the patient's eyes. The two cameras produce real-time video streams throughout the stimulation session and transmit them to the \textbf{CM}, allowing clinicians to continuously monitor the patient's eye opening, gaze direction, and fixation without interrupting the procedure. This solution has not been implemented in the cardboard headset due to a conflict with the required smartphone brightness. \\

\item Control Module 

The \textbf{CM} supports four primary functions: i) connecting to the \textbf{SM} using a pairing code generated at startup; ii) monitoring, in real-time, both the visual stimulus currently being preseted to the patient and, when using the modified Oculus Quest 2 headset, the video streams captured from the added eye-pointed cameras; iii) controlling stimulus playback, and iv) stimuli creation and editing. The application provides a comprehensive set of clinically relevant visual stimuli: 
\begin{itemize}
    \item \textbf{Configurable flashing lights.} Independent flash and background colors to play conventional IPS can be selected, which enables evaluating the photoprovocative effects of color and color-contrast. 
    \item \textbf{Photoprovocative patterns} \cite{Graciela2024,Wong1993}. Such patterns can be played either statically or with configurable oscillatory motions, such as translation, rotation, or zoom bouncing.
    \item Several \textbf{videoclips} which have been associated with massive epileptic seizure events, such as the Electric Soldier Porygon scene of the Pokémon's TV cartoon \cite{Ishida1998}, or the 2012 London Olympic Games TV advertisement \cite{Reuters2007}. VR-IPS enables the controlled display of real-world visual triggers under clinical conditions.
\end{itemize}

\end{enumerate}


\item Illustrative examples

To illustrate the application and its capabilities, Fig.\ref{fig:app} presents two different monitor layouts. The upper layout shows both modules on a single screen for demonstration purposes. On the top left, the \textbf{CM} provides real-time monitoring of the patient's eyes and, at the same time, the current stimulus the patient is exposed to.  On the bottom left, the playback controls that allow managing the stimulation procedure (go to previous onset, stop, pause, go to next onset, and loop, respectively from left to right), and four tabs to access the four available stimuli categories (flashing lights, static patterns, dynamic patterns, and videoclips, respectively from left to right). On the top right, the \textbf{SM} screens the selected stimulus. 

The lower layout in Fig.\ref{fig:app} presents a more realistic setup. The \textbf{CM} has been designed to share the screen with the EEG visualization software of the neurophysiology service, so there is no need for a second monitor. The \textbf{CM} automatically rearranges to occupy minimal screen space while preserving easy access to all of its components, thus maximizing the screen area available for the EEG visualization software. Meanwhile, the \textbf{SM} runs independently on the chosen stimulation device.

\begin{figure}[h!]
    \centering
    \includegraphics[width=0.85\linewidth]{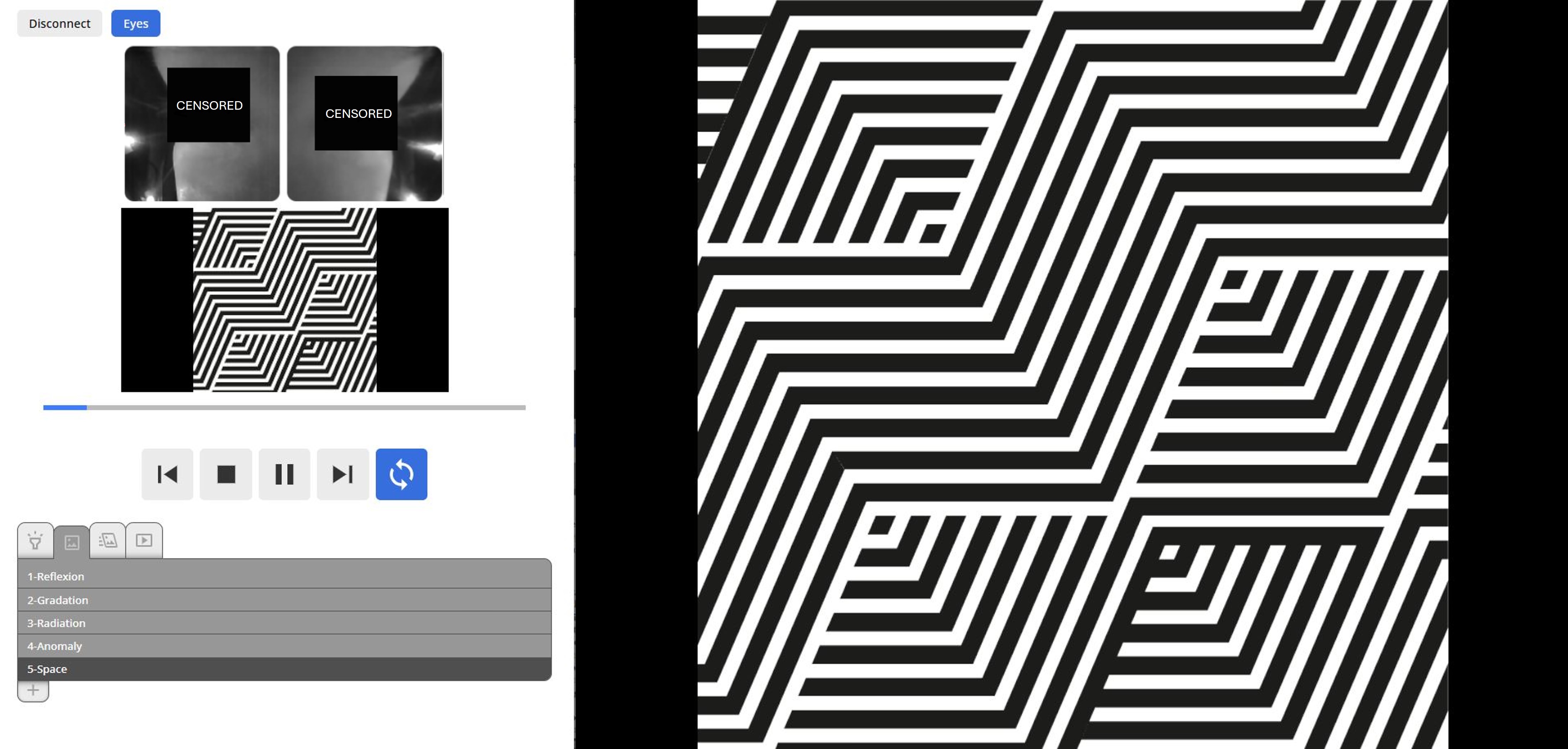}\\
    \vspace{5mm}
    \includegraphics[width=0.85\linewidth]{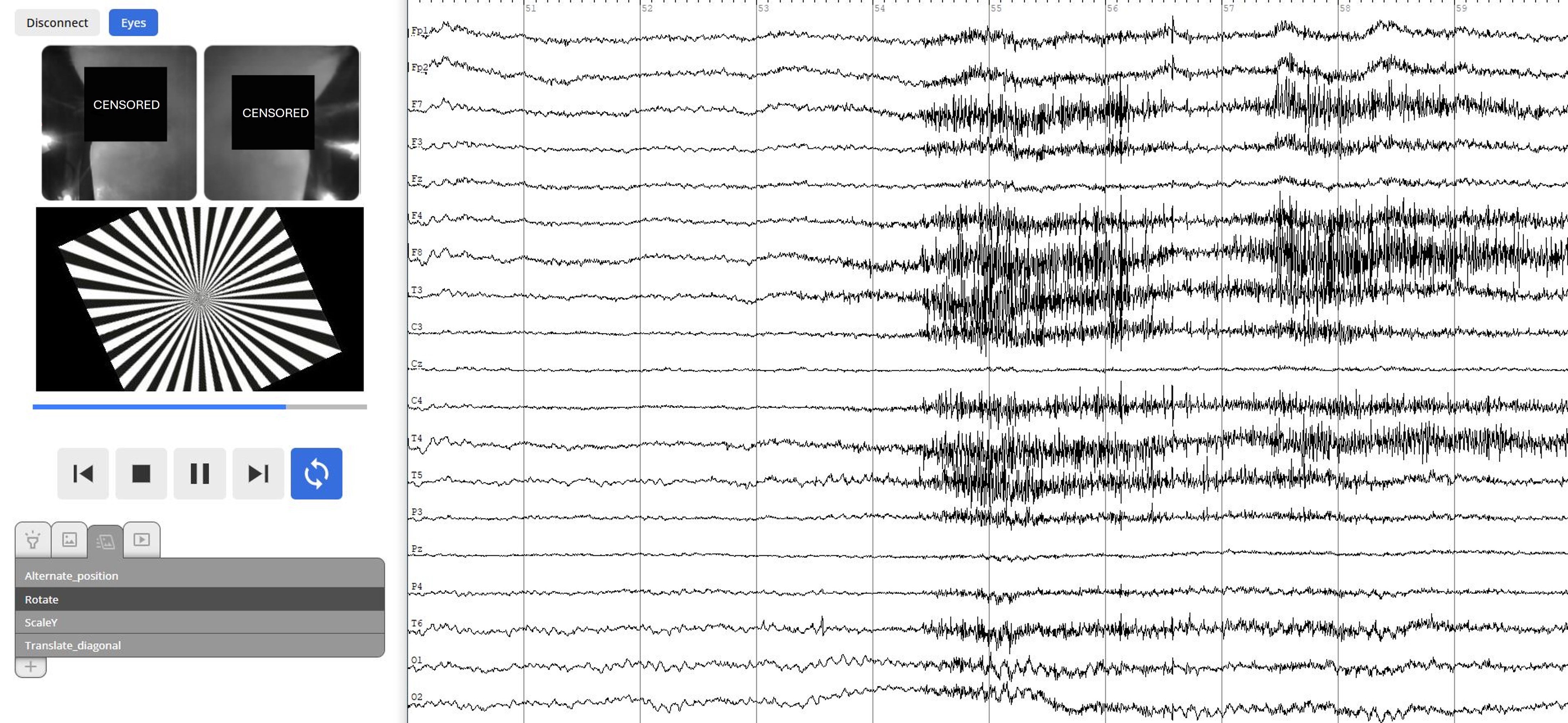}
    \caption{On the \textbf{upper half}: the two modules that comprise the application, with the \textbf{CM} shown on the left, and the \textbf{SM} on the right. On the \textbf{bottom half}: a realistic clinical layout that combines the \textbf{CM} and the EEG real-time visualization.}
    \label{fig:app}
\end{figure}

Stimulus creation and editing are performed through the configuration interface shown in Fig.\ref{fig:config}. Each stimulus can be assigned a descriptive name and configured according to its type. For flashing-light stimuli, parameters such as sequence of stimulation frequencies, stimulation and resting times, and foreground and background colors can be adjusted. For dynamic patterns, clinicians can select the motion type and configure the speed, direction, and visual pattern to use. Existing stimuli can also be easily duplicated or deleted, facilitating the creation of different customized versions. \\

\begin{figure}[h!]
    \centering
    \includegraphics[width=0.85\linewidth]{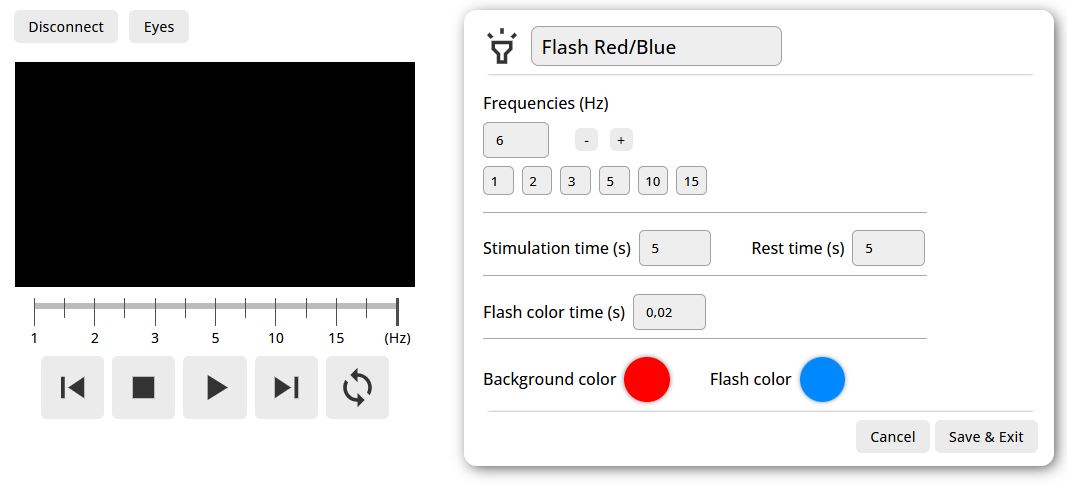}
    \caption{Configuration window for a flashing stimulus.}
    \label{fig:config}
\end{figure}


\item Impact 

VR-IPS has the potential to generate a meaningful impact in both clinical practice and research by expanding the range of visual stimulation paradigms available for the diagnosis of photosensitivity. 

From a clinical perspective, the app provides clinicians with a broader repertoire of photic and visual stimuli than those available through conventional IPS. By supporting both immersive VR and monitor-based stimulation, it enables the controlled presentation of visual conditions that are difficult or impossible to play with the traditional IPS. These capabilities may contribute to a more comprehensive evaluation of patients undergoing photosensitivity assessment and support the investigation of cases in which conventional IPS does not elicit PPR activity.

For clinicians, the system offers a flexible and configurable platform that can be adapted to different diagnostic protocols and research requirements. In addition to playing the conventional IPS, it enables the creation of customized stimulation sequences, which may leverage the evaluation of novel stimulation paradigms. Thus, it constitutes a much more powerful stimulation tool than conventional IPS, expanding the range of diagnostic tools available during clinical examinations.

From a research point of view, VR-IPS provides a versatile experimental platform for investigating photosensitivity and evaluating novel visual stimulation protocols. Its software-based architecture enables rapid implementation and deployment of new stimulation paradigms without requiring modifications to dedicated stimulation hardware. This flexibility facilitates systematic comparisons between conventional and a wide range of new stimuli, including immersive virtual reality environments, and promotes collaboration between clinicians and engineers. Consequently, the platform provides a foundation for future studies aimed at improving the understanding and clinical assessment of photosensitivity. \\

\item Conclusions 

VR-IPS extends conventional white-flashing-light stimulation for photosensitivity diagnosis by providing a flexible, software-based application capable of delivering a broad range of configurable visual stimulation paradigms on both conventional monitors and immersive VR devices. Developed in close collaboration with expert clinical neurophysiologists from Cabueñes University Hospital, the system was evaluated in their service. It provides immersive stimulation capabilities, an intuitive control interface, and extensive configuration options.

Future work will focus primarily on the integration and synchronization of the application with the hospital's EEG acquisition software to automatically signal stimulation events within EEG recordings. Improvements to the integrated eye-monitoring system are also planned, including repositioning the embedded cameras to provide a more frontal view of the patient's eyes. Additional developments include immersive 3D photoprovocative scenarios, configurable stimulation sequences tailored to individual patients, and a manual stimulation mode that allows clinicians to modify stimulation parameters on the go.

\end{enumerate}


\section*{Acknowledgements}
\label{}

This research has been funded by the Spanish Research Agency --grant PID2023-146257OB-I00--. Also, by Principado de Asturias, grant IDE/2024/000734, and by the Council of Gijón through the University Institute of Industrial Technology of Asturias grants SV-24-GIJÓN-1-05 and SV-25-GIJÓN-1-14.

\section*{Author Contributions: CRediT}
\textbf{Fernando Moncada Martins:} Conceptualization, Investigation, Methodology, Project administration, Supervision, Writing – original draft, Writing – review and editing; \textbf{Daniel P{\'e}rez Pr{\'a}danos:} Software, Visualization; \textbf{Angel Rio-Alvarez:} Supervision; \textbf{MG Cano-Celestino:} Conceptualization, Investigation; \textbf{Mar{\'i}a Antonia Guti{\'e}rrez:} Validation, Writing – review and editing; \textbf{Pablo Calvo Calleja:} Validation, Writing – review and editing; \textbf{D. Kasteleijn-Nolst Trenite:} Validation, Writing – review and editing; \textbf{V{\'i}ctor M. Gonz{\'a}lez:} Conceptualization, Funding acquisition, Methodology, Project administration, Supervision, Writing – original draft, Writing – review and editing.



\section*{Current executable software version}
\label{}
The application is available and ready to execute in \url{https://visir.edv.uniovi.es}

\end{document}